\documentclass{aa}
\usepackage{graphicx}
\usepackage{txfonts}
\usepackage{natbib}
\usepackage[figuresright]{rotating}
\usepackage[a4paper,breaklinks]{hyperref}
\idline{1}{5}
\begin{document}
\def\teff{$T\rm_{eff}$ }
\def\kms {\,$\mathrm{km\, s^{-1}}$ }
\def\kmss {\,$\mathrm{km\, s^{-1}}$}
\def\ms {$\mathrm{m\, s^{-1}}$ }

\newcommand{\Teff}{\ensuremath{T_\mathrm{eff}}}
\newcommand{\g}{\ensuremath{g}}
\newcommand{\gf}{\ensuremath{gf}}
\newcommand{\loggf}{\ensuremath{\log\gf}}
\newcommand{\glog}{\ensuremath{\log\g}}
\newcommand{\pun}[1]{\,#1}
\newcommand{\cobold}{\ensuremath{\mathrm{CO}^5\mathrm{BOLD}}}
\newcommand{\linfor}{Linfor3D}
\newcommand{\xx}{\ensuremath{\mathrm{1D}_{\mathrm{LHD}}}}
\newcommand{\punms}{\mbox{\rm\,m\,s$^{-1}$}}
\newcommand{\punkms}{\mbox{\rm\,km\,s$^{-1}$}}
\newcommand{\abuhe}{\mbox{Y}}
\newcommand{\grav}{\ensuremath{g}}
\newcommand{\mlp}{\ensuremath{\alpha_{\mathrm{MLT}}}}
\newcommand{\mlpcm}{\ensuremath{\alpha_{\mathrm{CMT}}}}
\newcommand{\moh}{\ensuremath{[\mathrm{M/H}]}}
\newcommand{\senv}{\ensuremath{\mathrm{s}_{\mathrm{env}}}}
\newcommand{\shelio}{\ensuremath{\mathrm{s}_{\mathrm{helio}}}}
\newcommand{\smin}{\ensuremath{\mathrm{s}_{\mathrm{min}}}}
\newcommand{\spun}{\ensuremath{\mathrm{s}_0}}
\newcommand{\sstar}{\ensuremath{\mathrm{s}^\ast}}
\newcommand{\tauross}{\ensuremath{\tau_{\mathrm{ross}}}}
\newcommand{\ttaurelation}{\mbox{T$(\tau$)-relation}}
\newcommand{\Ysurf}{\ensuremath{\mathrm{Y}_{\mathrm{surf}}}}
\newcommand{\mD}{\ensuremath{\left\langle\mathrm{3D}\right\rangle}}

\newcommand{\draftflag}{false}

\newcommand{\beq}{\begin{equation}}
\newcommand{\eeq}{\end{equation}}
\newcommand{\pdx}[2]{\frac{\partial #1}{\partial #2}}
\newcommand{\pdf}[2]{\frac{\partial}{\partial #2}\left( #1 \right)}

\newcommand{\var}[1]{{\ensuremath{\sigma^2_{#1}}}}
\newcommand{\sig}[1]{{\ensuremath{\sigma_{#1}}}}
\newcommand{\cov}[2]{{\ensuremath{\mathrm{C}\left[#1,#2\right]}}}
\newcommand{\xtmean}[1]{\ensuremath{\left\langle #1\right\rangle}}

\newcommand{\eref}[1]{\mbox{(\ref{#1})}}

\newcommand{\Vact}{\ensuremath{\nabla}}
\newcommand{\Vad}{\ensuremath{\nabla_{\mathrm{ad}}}}
\newcommand{\Veddy}{\ensuremath{\nabla_{\mathrm{e}}}}
\newcommand{\Vrad}{\ensuremath{\nabla_{\mathrm{rad}}}}
\newcommand{\Vraddiff}{\ensuremath{\nabla_{\mathrm{rad,diff}}}}
\newcommand{\cp}{\ensuremath{c_{\mathrm{p}}}}
\newcommand{\taueddy}{\ensuremath{\tau_{\mathrm{e}}}}
\newcommand{\vconv}{\ensuremath{v_{\mathrm{c}}}}
\newcommand{\Fconv}{\ensuremath{F_{\mathrm{c}}}}
\newcommand{\lmix}{\ensuremath{\Lambda}}
\newcommand{\Hp}{\ensuremath{H_{\mathrm{P}}}}
\newcommand{\Hptop}{\ensuremath{H_{\mathrm{P,top}}}}
\newcommand{\COBOLD}{{\sc CO$^5$BOLD}}

\newcommand{\changed}{}

\newcommand{\I}{\ensuremath{I}}
\newcommand{\Irot}{\ensuremath{\tilde{I}}}
\newcommand{\F}{\ensuremath{F}}
\newcommand{\Frot}{\ensuremath{\tilde{F}}}
\newcommand{\vsini}{\ensuremath{V\sin(i)}}
\newcommand{\vvsini}{\ensuremath{V^2\sin^2(i)}}
\newcommand{\vsinimu}{\ensuremath{\tilde{v}}}
\newcommand{\rotint}{\ensuremath{\int^{+\vsinimu}_{-\vsinimu}\!\!d\xi\,}}
\newcommand{\imu}{\ensuremath{m}}
\newcommand{\imupone}{\ensuremath{{m+1}}}
\newcommand{\nmu}{\ensuremath{N_\mu}}
\newcommand{\msum}[1]{\ensuremath{\sum_{#1=1}^{\nmu}}}
\newcommand{\wmu}{\ensuremath{w_\imu}}

\newcommand{\tchar}{\ensuremath{t_\mathrm{c}}}
\newcommand{\Nt}{\ensuremath{N_\mathrm{t}}}

\title{The photospheric solar oxygen project:\thanks{Based on observations collected at ESO Paranal Observatory, Programme 182.D-5053(A)}}
\subtitle{II. Non-concordance of the oxygen abundance derived from two forbidden lines}

\author{
E. Caffau\thanks{Gliese Fellow}\inst{1,2}\and
H.-G. Ludwig  \inst{1,2}\and
J.-M. Malherbe \inst{3}\and
P. Bonifacio  \inst{2}\and
M. Steffen   \inst{4,2}\and
L. Monaco     \inst{5}
}

\institute{
Zentrum f\"ur Astronomie der Universit\"at Heidelberg, Landessternwarte, 
K\"onigstuhl 12, 69117 Heidelberg, Germany
\and
GEPI, Observatoire de Paris, CNRS, Universit\'e Paris Diderot, Place
Jules Janssen, 92190
Meudon, France
\and
LESIA, Observatoire de Paris, CNRS, Universit\'e Paris Diderot, Place
Jules Janssen, 92190
Meudon, France
\and
Leibniz-Institut f\"ur Astrophysik Potsdam (AIP), An der Sternwarte 16, 14482 Potsdam, Germany
\and
Leibniz-Institut f\"ur Astrophysik Potsdam, An der Sternwarte 16, 
D-14482 Potsdam, Germany
\and
European Southern Observatory, Casilla 19001, Santiago, Chile
}
\authorrunning{Caffau et al.}
\titlerunning{Solar oxygen project II}
\offprints{}
\date{Received 12/03/2013; Accepted 26/04/2013}

\abstract
{In the Sun, the two forbidden [OI] lines at 630 and 636 nm were previously found 
to provide discrepant oxygen abundances.}
{We  investigate whether this discrepancy is peculiar
to the Sun or whether it is also observed in other stars.}
{We make use of high-resolution, high signal-to-noise ratio
spectra of four dwarf to turn-off stars, five giant stars, and one sub-giant star
observed with THEMIS, HARPS, and UVES to investigate
the coherence of the two lines.  
}
{The two lines provide oxygen abundances that are
consistent, within observational errors, in all the
giant stars examined by us. On the other hand, for 
the two dwarf stars for which a measurement was
possible, for Procyon, and for the sub-giant star Capella,
the 636\,nm line provides systematically higher
oxygen abundances, as already seen for the Sun.
}
{The only two possible reasons for the discrepancy
are a serious error in the oscillator strength
of the \ion{Ni}{I} line blending the
630 nm line or the presence of an unknown blend
in the 636 nm line, which makes the feature
stronger. The CN lines blending the
636 nm line cannot be responsible for the discrepancy. 
The \ion{Ca}{i} autoionisation line, on the 
red wing of which the 636 nm line is formed,
is not well modelled by our synthetic spectra. 
However, a better reproduction of this line would
result in even higher abundances
from the 636\,nm, thus increasing the discrepancy.
}

\maketitle


\section{Introduction}

We investigate the discrepancy in the oxygen abundance 
derived from the two forbidden oxygen lines at 630.0 and 636.3\,nm.
This discrepancy in the solar photosphere was highlighted by our analysis
\citep{oxy}. 
We wish to establish if this disagreement is peculiar to the Sun or 
if it is shared by other stars.
In this context, we are not interested in the oxygen abundance itself,
but in the agreement of the two [OI] forbidden lines with respect to
the abundance determination.
In \citet{oxy} we investigated four solar atlases, and for all of them
we found that the oxygen abundance derived from the resonance line
at 630\,nm is about 0.1\,dex lower than the abundance derived from the
subordinate line at 636\,nm.
The [OI] lines at 630 and 636\,nm are blended, with a \ion{Ni}{i} line and
some weak CN lines, respectively.
These CN lines are mainly on the red wing of the blend: they are negligible
in solar temperature stars but become visible in cooler stars.
The subordinate line at 636\,nm also lies on the red wing of a \ion{Ca}{i}
auto-ionisation line.
Generally, we can envisage several explanations for the disagreement:
\begin{itemize}
\item the oscillator strength of one/both oxygen lines is in error 
\item the oscillator strength(s) of one/some blending line(s) is/are incorrect
\item there are unknown lines blending one/both [OI] lines
\item the line profile of the \ion{Ca}{i} auto-ionisation line is not modelled
with sufficient precision
\item there are some unknown blends making the 636\,nm feature stronger
\item hydrodynamical effects
\item non-local thermodynamical equilibrium effects.
\end{itemize}
We already investigated the last of these explanations in \citet{oxy} and feel confident 
about excluding it. The level populations of the [OI] lines are close to local thermodynamical equilibrium.
For CN and \ion{Ni}{i} lines there is no investigation of the deviation from
local thermodynamical equilibrium (NLTE), but we do not think
NLTE can be responsible for this disagreement. 
The \loggf\ values for the two [OI] lines are of good quality \citep{SZ} 
and come from the same computation. Thus 
we expect that, if any systematic error affects the calculation, 
it affects both lines
in the same way and both lines should scale in the same way.
We discard errors in the oscillator strengths of the oxygen
lines as a possible explanation for the discrepancy. 

In the following we describe the observations and analysis
of four dwarf, one sub-giant, and five giant stars. 
Since the different blending components have a different
response to effective temperature and gravity, we can make
some progress discarding some of the possible explanations.

A third forbidden oxygen line at 557.7\,nm is observable in the solar spectrum.
However, the oxygen contribution to the feature is minor with respect to the
contribution of C$_2$, for which only uncertain molecular data are available
\citep[see][]{melendezoi08}.

\begin{table}
\caption{[OI] line parameters.}
\label{oxyline}
\begin{center}
{
\begin{tabular}{rrrr}
\hline
\noalign{\smallskip}
 Wavelength & Transition & E$_{\rm low}$ & \loggf  \\
 nm         &            & eV\\
\noalign{\smallskip}
\hline
\noalign{\smallskip}
 630.0304  & $^3$P$-^1$D &  0.000 &  --9.717\\
 636.3776  & $^3$P$-^1$D &  0.020 & --10.185\\
\noalign{\smallskip}
\hline
\end{tabular}
}
\end{center}
\end{table}


\section{Atomic data and model atmospheres}

For the [OI] lines we adopted \loggf\ from \citet{SZ}
listed in Table\,\ref{oxyline}.
For the \ion{Ni}{i} blending line we adopted the \loggf = --2.11
from \citet{Johansson} and an isotopic structure as described in
\citet{oxy}.
For CN molecules we used the line list of Kurucz\footnote{http://kurucz.harvard.edu/}.
We also tested the CN atomic data kindly provided by Bertrand Plez.
There are differences between the two line lists overall for the 636\,nm line,
but this more updated list has the effect of increasing the solar abundance of
oxygen derived from this line, aggravating the disagreement 
between the two lines as A(O) indicators. 

For each star we computed an ATLAS\,9 or ATLAS\,12 model,
running the Linux version of ATLAS \citep{kurucz05,sbordone04,sbordone05}.
In the cases for which an opacity distribution functions (ODF) 
with similar parameters was available,
we ran an ATLAS\,9 model; otherwise we ran an ATLAS\,12.
The line profiles were computed with SYNTHE \citep{synte93,kurucz05}
in its Linux version \citep{sbordone04,sbordone05}.


\section{Observations}

We obtained high-resolution, high signal-to-noise (S/N) ratio
spectra with THEMIS \citep{themis} and HARPS  (see below)
for five stars.
We complemented this data with high-quality 
spectra for the other four stars obtained with either HARPS or UVES \citep{Dekker},
which were retrieved from the ESO archive\footnote{archive.eso.org} or
the UVES-POP database \citep{uvespop}. For three of the stars
observed with THEMIS we also retrieved UVES spectra from the 
UVES-POP database.

\subsection{THEMIS}

Observations with the French telescope THEMIS (INSU/CNRS), which is
located at Observatorio del Teide, Tenerife, Canary Islands, were
obtained in April and September 2011. THEMIS is
an altazimuthal one-meter telescope with primary focus at f/17 and secondary
focus at f/60. A tip-tilt mirror is located at the secondary
focus for image stabilisation and feature tracking. The slit of
the spectrograph has an aperture of 0.5 arc sec. The spectrograph
operates with two gratings: a predisperser and a disperser.
The dispersion is 5 mm for 0.1 nm wavelength at 630\,nm.
The resolution of the spectrograph in the red part of the spectrum is
R = 300\,000. Data were registered using a cooled EMCCD device (electron
multiplication) with high gain
(100) for stellar spectra. The spectral pixel on the EMCCD device
was 12\,m\AA\ in the spectral direction and 0.2 arc sec in the spatial
direction. The typical exposure time was one hour for the stellar spectra 
(S/N ratio $>$100, depending on the magnitude).

\subsection{HARPS}
Stellar spectra observed with the High Accuracy Radial velocity Planet 
Searcher\citep[HARPS;][]{2003Msngr.114...20M} spectrograph 
mounted at the 3.6\,m
telescope at the La\,Silla ESO observatory were also obtained from the
pipeline processed data
archive\footnote{http://archive.eso.org/wdb/wdb/eso/repro/form}. HARPS
is a fibre-fed, cross-dispersed echelle spectrograph, which provides a
resolution R=115.000 with a fixed format covering the 380-690\,nm
spectral range. The HARPS pipeline delivers science-quality products. In
particular, master calibrations are created (master-bias and master-flat
fields), and data are bias subtracted and divided by the flat field by
using the calibrations closest in time to the science observations.
Spectral orders are located and extracted, and the wavelength
solution obtained using a Thorium-Argon calibration lamp is then applied.
We refere to the documentation of the data reduction pipeline for further
information\footnote{http://www.eso.org/sci/facilities/lasilla/instruments/harps/doc/index.html}.

\subsubsection{HARPS observations}

We observed HD\,26297 on the night of September 19th, 2008. Two 1\,hr plus
one 50\,min exposure were collected with air mass (AM) varying in the range
1.5-1.1. The seeing conditions were variable between  1-1.5\,arcsec, and we
obtained a S/N ratio of about 300 at 630\,nm in the single frames, which converted
to a S/N$\simeq 450$ of the median combined, final spectrum.

\subsubsection{HARPS archival data}

Archive HARPS spectra for HD\,30562 were retrieved from the ESO archive.
Observations were taken on November 17th, 2011 at AM$\simeq$1.3 and with seeing
conditions below 1 arcsec. Three 300\,s spectra were retrieved, yielding a S/N
ratio of the median spectrum of about 280 at 630\,nm.

For $\alpha$\,Cen A 
we used a sequence of 12 spectra, each of 2\,s exposure,
observed on 10/06/2006; 
the spectra were co-added.
For $\alpha$\,Cen   B
we co-added (after shifting to rest wavelength)
182 exposures, obtained between 10/06/2006 and 30/03/2009.
The exposures were either of 5\,s or 10\,s duration.
The resulting S/N ratio is better than 500 for both co-added
spectra.

\subsection{UVES}

The UVES spectra for Aldebaran, Arcturus, Procyon,
and HD 152786 were downloaded from the
UVES-POP database 
\citep{uvespop}\footnote{http://www.eso.org/sci/observing/tools/uvespop.html.html}.
These spectra have all been acquired with a 0\farcs{5} slit
that provides a resolution of about 80\,000.

\begin{table*}
\caption{Results}
\label{results}
\begin{center}
\begin{tabular}{llrrrrrrrrrr}
\hline
\noalign{\smallskip}
Star & Instr. & \teff & \glog & [Fe/H] & [C/H] & [N/H] & [Ni/H] & A(O)$_{630}$ & A(O)$_{636}$ & EW$^1_{630}$ & EW$^1_{636}$\\
\noalign{\smallskip}
\hline
\noalign{\smallskip}
Aldebaran      & THEMIS & 3891 & 1.20 & -0.15 & 8.30 & 8.16 & 6.08 & 8.74 & 8.76 & 7.5 & 4.6 \\ 
Arcturus       & THEMIS & 4300 & 1.50 & -0.50 & 8.00 & 7.36 & 5.75 & 8.57 & 8.54 & 7.1 & 3.2 \\ 
Pollux         & THEMIS & 4750 & 2.74 & 0.00  & 8.52 & 7.92 & 6.25 & 8.78 & 8.81 & 3.0 & 1.3 \\ 
HD\,152786     & UVES   & 4350 & 1.50 & -0.40 & 8.10 & 7.46 & 5.85 & 8.87 & 8.90 & 8.4 & 4.3 \\ 
HD\,26297      & HARPS  & 4400 & 1.10 & -1.87 & 7.17 & 5.51 & 4.06 & 7.51 & 7.52 & 4.0 & 1.3 \\
Capella        & THEMIS & 5270 & 3.05 & -0.37 & 8.13 & 7.49 & 5.88 & 8.34 & 8.62 & 1.3 & 3.7 \\ 
HD\,30562      & HARPS  & 5876 & 4.00 & +0.19 & 8.71 & 8.11 & 6.64 & $\le 8.62$& $\ge 8.62$ & 5.0 & 0.3\\ 
$\alpha$\,Cen A & HARPS  & 5824 & 4.34 & +0.24 & 8.76 & 8.16 & 6.49 & 8.82 & 8.93 & 0.8 & 0.3 \\ 
$\alpha$ Cen B & HARPS  & 5223 & 4.44 & +0.25 & 8.77 & 8.17 & 6.50 &  -  & - & 0.5 & 0.3 \\ 
Sun            &        & 5780 & 4.44 & +0.00 & 8.50 & 7.86 & 6.25 & 8.69 & 8.81 & 0.47 & 0.14 \\
Procyon        & UVES   & 6500 & 4.00 & 0.00  &      &      &      & $< 8.56$ & $> 8.56$ & & \\   
\noalign{\smallskip}
\hline
\end{tabular}
\end{center}
$^1$ EWs take into account the contributions of oxygen and all blends;
they are not used for the abundance analysis, but give an order of magnitude of the strength of the lines.
\end{table*}


\section{Data analysis}

\subsection{Sun} \label{sun}

We already analysed the two solar [OI] lines in \citet{oxy}, where
we derived the abundance from equivalent width (EW) measurements
and used as model a \cobold\ \citep{freytag02,freytag12} solar model.
The oxygen abundances derived from the two forbidden lines at 630 and 636\,nm
in \citet{oxy} are $8.68\pm 0.15$ and $8.78\pm 0.12$, respectively.
However, there is no unanimous opinion about the disagreement 
of the oxygen abundance derived from the two [OI] lines.
\citet{asplund04} find a very good agreement between the two oxygen abundances 
derived from the two [OI] lines (A(O)=8.69 and 8.67, respectively).
More recently the oxygen derived from the two [OI] lines
by the group led by Asplund is quoted as
A(O) of 8.64 and 8.70, respectively \citep{stasinska12}.

In the present work we are not interested in the solar oxygen abundance
as such, but in the disagreement between the A(O) derived from the
two forbidden lines.
In preparation is a detailed study of the two forbidden lines in the solar photosphere,
with spectra observed especially for this purpose, as well as a discussion
of the results derived from other groups.

We here approach the solar spectra in the same way as we do for the
other stars to be consistent with the analysis of the other
stars when we compare them, with the just difference that we consider only
two disc-centre spectra. 
We already know from \citet{oxy} that
disc-centre and integrated-disc spectra give consistent results
with respect to the disagreement between the oxygen abundance derived
from the two [OI] lines; we prefer to do the analysis
with disc-centre spectra because the full width half
maximum of the lines is smaller than in the integrated disc spectra.
We analysed two disc-centre solar spectra: \citet{neckelobs}
absolutely calibrated Fourier transform spectra obtained at Kitt Peak, and
\citet{delbouille} observed at 
Jungfraujoch\footnote{http://bass2000.obspm.fr/solar\_spect.php}.

As solar model atmospheres we used
the model obtained by averaging each 3D snapshot of the \cobold\
solar model, \mD, used in \citet{oxy} over surfaces of equal
(Rosseland) optical depth.
For the blending lines, the adopted abundances of the relevant elements are
A(C)=8.50 and A(N)=7.86 from \cite{abbosun} and A(Ni)=6.25 from \cite{grevesse98}.
In this analysis, as well as in \citet{oxy}, the abundance
from the subordinate line comes out more than 0.1\,dex higher
than from the resonance line when analysing the Kitt Peak solar atlas (see Table\,\ref{oxyline}).
In the case of the Jungfraujoch spectrum, the disagreement from the two lines is 0.08\,dex.

According to a SYNTHE simulation, by using as input the \mD\ model,
the Ni contribution to the total EW of the 630\pun{nm} blend is
about 36.7\%, 31.7\%, and 27.0\% if A(O) is of 8.56, 8.66, and 8.76, respectively.
The CN contribution to the total EW of the 636\pun{nm} line is
18.8\%, 15.6\%, and 12.9\%, if A(O) is of 8.56, 8.66, and 8.76, respectively. 

The best fit (lowest reduced $\chi ^2$ as described in \citealt{fit03}) 
of the subordinate line is obtained with a 
broadening larger (by about 0.5\kms)
than the one required for the best fit of the forbidden line at 630\,nm.
A change of 0.5\kms\ in the broadening produces a change in the
oxygen abundance of 0.02\,dex from the 636\,nm line.
In conclusion, with the broadening fixed at the value derived from the best fit 
of the line at 630\,nm, the fit of the subordinate line is not as well reproduced
as the [OI] 630\,nm line.

An increase of 0.2\,dex in the oscillator strengths of the CN lines
blending the subordinate line is not enough to decrease significantly 
the derived oxygen abundance.
It decreases the A(O) by 0.06\,dex, which is the right direction, but not enough.

The shape of the auto-ionisation \ion{Ca}{i} line is not well reproduced
by the synthetic spectra. The Ca line seems somewhat deeper in the observed
spectrum. We increased the Ca abundance to improve the agreement
between observed and synthetic spectra. 
An increase of 0.1\,dex in A(Ca) improves the all-in-all
comparison of the range, but increases the oxygen abundance by 0.02\,dex.
This difference is negligible and anyway in the wrong direction to improve
the agreement between the two lines.
We also shifted a high-excitation \ion{Fe}{i} line so that it blends the [OI] 636\,nm line, and the presence
of this line decreases A(O) derived from this line by 0.05\,dex (see Section\,5 for details).
A decrease in the Ni abundance of 0.1\,dex implies an increase of A(O) by about 0.04\,dex.

\subsection{Aldebaran}

\begin{figure}
\resizebox{\hsize}{!}{\includegraphics[clip=true,angle=0]{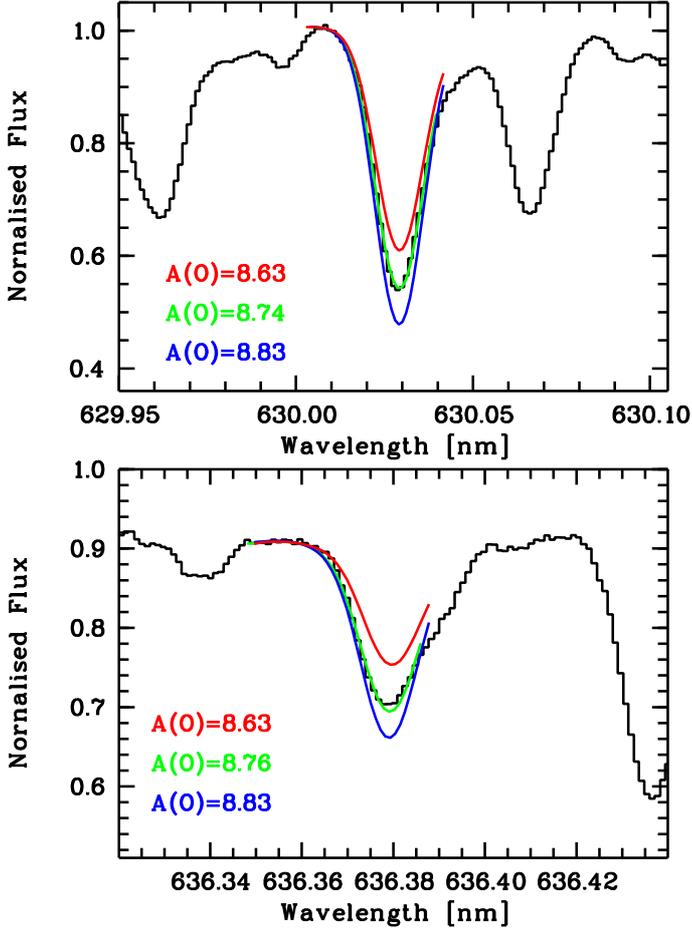}}
\caption{Spectra of Aldebaran observed with THEMIS (solid black) overimposed to
the best fit (solid green) for the
630\pun{nm} and the 636\pun{nm} lines.
In red and blue are two synthetic spectra that differ by about 0.1\,dex in the O abundance from the best fit.
}
\label{aldebaran}
\end{figure}

We analysed a spectrum observed with THEMIS and two spectra observed with 
UVES-VLT that we downloaded from the UVES-POP database \citep{uvespop}.
The stellar parameters \teff/\glog/[M/H] (3891/1.2/--0.15) 
and the abundances of the relevant
elements ([C/Fe]=--0.20, [N/Fe]=+0.30) are from \citet{melendez08}.
We computed an ATLAS\,9 model with parameters 3891/1.2/0.0
and adjusted the abundances in the spectral-synthesis computation.
We scale Ni as Fe.
With a line-profile fitting of the two [OI] lines and an ATLAS+SYNTHE grid,
we derive the oxygen abundances shown in Table\,\ref{results}.
The agreement between the observations we have from the two spectrographs is 
good, but the data are not of the quality to derive A(O)
from such complicated spectra. The uncertainty related to the continuum
placement and the broadening is large (on the order of 0.04\,dex).
The oxygen abundance from the [OI] 630\,nm line is in close agreement with
the abundance derived from the subordinate line,
but there are several uncertainties due to
a feature on the red side of the 636.6\,nm line that is not reproduced
in the synthetic spectrum.
The best fits are shown in Fig.\,\ref{aldebaran}, but as it can be seen
in the plot, the fit on the subordinate line does not incorporate the
red wing.

According to an ATLAS+SYNTHE simulation, 
the Ni contribution to the complete EW of the 
630\pun{nm} blend is negligible. This is also because the
630\,nm line is saturated (EW of about 7.5\,pm) and an EW of about 0.6\,pm
of the Ni line has no effect on the A(O) determination. 
As expected, a change of 0.1\,dex in the Ni abundance
does not affect the oxygen abundance derived from the [OI] 630\,nm line.
The CN contribution to the complete EW in the 636\pun{nm} line is
still small, but somewhat larger than the Ni contribution in the
O+Ni blend at 630\,nm (about 16\% in EW, but one should take into
account the saturation effects). 
The largely prevalent contribution in both features is due to oxygen.

\subsection{Arcturus}

We analysed four spectra of the 630.0\,nm [OI] line and two
of the 636.3\,nm [OI] line, observed with THEMIS, and a spectrum observed with 
UVES-VLT that we downloaded from the UVES-POP database \citep{uvespop}.
The stellar parameters that we adopted are ${\rm T_{\rm eff}}$/\glog/[M/H] = 4300/1.5/--0.5,
and [$\alpha$/Fe]=0.0.
The oxygen abundance that we derive
from line profile fitting of the two [OI] lines with an ATLAS+SYNTHE grid
are given in Table\,\ref{results}.
Both the agreement between the two spectrographs and the agreement
between the two [OI] lines are good.
The oxygen abundance of the 630\,nm line is insensitive to the Ni abundance
because the contribution from Ni in the O+Ni blend is negligible.
According to a SYNTHE simulation, when an input ATLAS model is used,
the Ni contribution to the complete EW of the 630\pun{nm} blend is small (6.4\% in EW) and
the CN contribution to the complete EW in the 636\pun{nm} line is similar (6.9\%). 

\begin{figure}
\resizebox{\hsize}{!}{\includegraphics[clip=true,angle=0]{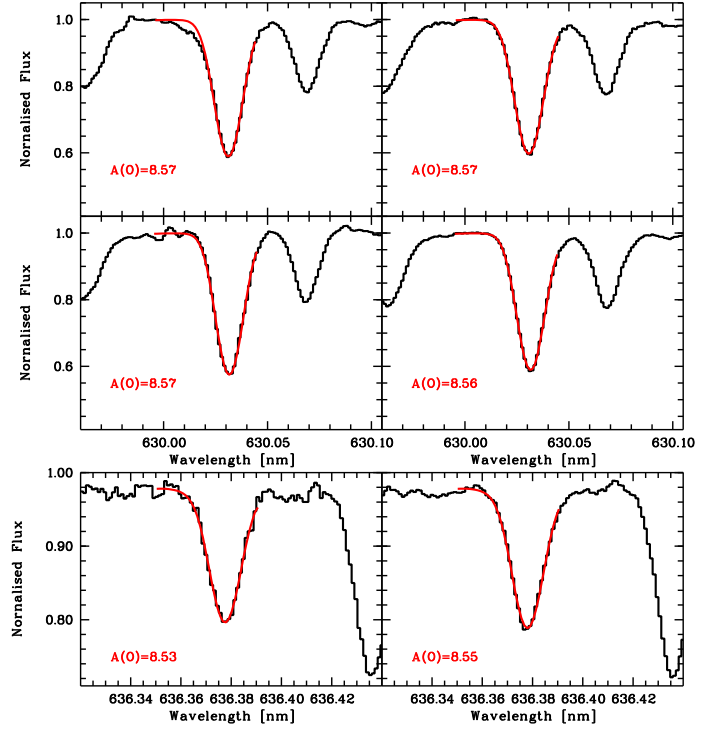}}
\caption{THEMIS observed spectra of Arcturus (solid black) and overimposed
the synthetic spectrum ATLAS+SYNTHE with the oxygen abundance obtained
from the best fit (solid red), for the
the 630\pun{nm} and the 636\pun{nm} lines.
}
\label{arcturus_fit}
\end{figure}

The abundance of the relevant elements is
A(C)=8.00, A(N)=7.36, and A(Ni)=5.75.
Adopting A(C)=8.06, A(N)=7.55 the agreement between the oxygen
abundance derived from the two [OI] lines is still good.
The fits are shown in Fig.\,\ref{arcturus_fit}

\subsection{Pollux}

\begin{figure}
\resizebox{\hsize}{!}{\includegraphics[clip=true,angle=0]{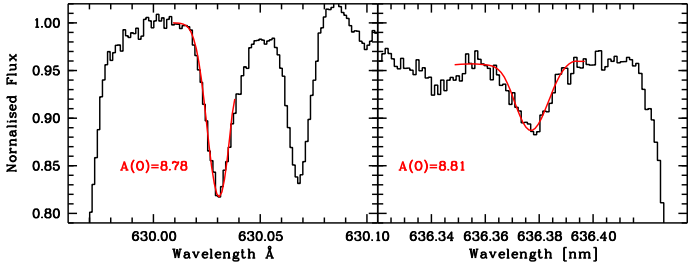}}
\caption{THEMIS observed spectra of Pollux (solid black) and overimposed
the best fit (solid red) for 
the 630\pun{nm} and the 636\pun{nm} lines.
}
\label{polluxfit}
\end{figure}

Pollux is another giant, 4750K/2.74/0.0 \citep{mallik98},
observed with THEMIS, for which we have a reasonable agreement 
between the oxygen derived from the two [OI] lines,
giving a difference of 0.03\,dex (see Fig.\,\ref{polluxfit} and Table\,\ref{results}).
The uncertainty in the continuum placement for the 636\, line is up to 0.05\,dex.
This is a solar metallicity giant, but the contribution
of the Ni line in the O+Ni blend is negligible.
We fitted the line profile with a grid of synthetic spectra containing
only the lines of O and Ni. The neglect of the other lines in the range
increases the O abundance by 0.01\,dex.
For a decrease of 0.1\,dex in the Ni abundance (from 6.25 to 6.15),
A(O) increases by less than 0.02\,dex.
The major contribution  in both blends is oxygen.

\subsection{HD 152786}

HD\,152786 has stellar parameters very similar to those Arcturus
(4350\pun{K}/1.5/--0.40). We retrieved spectra from the UVES-POP database
\citep{uvespop}.  We obtain satisfactory agreement between the abundance
derived from the two oxygen lines (see Table\,\ref{results}),
assuming A(C)=8.10,
A(N)=7.46, and A(Ni)=5.85.  
The Ni contribution in the blend at 630\,nm is very small.

\subsection{HD 26297}

We decided to analyse a metal-poor giant to understand if the
disagreement in A(O) seen in un-evolved stars can be related
to the strength of the lines. In fact, at fixed metallicity
the [OI] lines are much stronger in the spectra of giant stars.
We computed an ATLAS\,12 model with the parameters from \citet{burris00},
which are \teff of 4400\,K, gravity of 1.1, and [Fe/H]=--1.87.
We fixed C and N from \citet{gratton86} and
Ca and Ni from \citet{pilachowski96}.
The oxygen abundances derived from the two lines are in very good agreement
(see Table\,\ref{results}). 

\subsection{Capella}

The stellar parameters we adopted are from \citet{mcwilliam90} 
(\teff =5270\,K, \glog =3.05, [M/Fe]=--0.37).
The spectra are not of as good quality as for Aldebaran and Arcturus,
but this suboptimal quality cannot justify the disagreement in 
the oxygen abundance derived from the two
forbidden lines of almost 0.3\,dex, that is A(O) of 8.34 and 8.62
from the line at 630 and 636\,nm, respectively
(see Table\,\ref{results}) with A(C)=8.13, A(N)=7.49, and A(Ni)=5.88.
The model atmosphere used was with [$\alpha$/Fe]=0.
If we take the model atmosphere with [$\alpha$/Fe]=0.4,
we obtain A(O) of 8.46 and 8.76
from the line at 630 and 636\,nm, respectively.

According to ATLAS+SYNTHE simulations, when an input ATLAS model with [$\alpha$/Fe]=0 is used,
the Ni contribution to the complete EW of the 630\pun{nm} blend is
non-negligible (18.1\% in EW), a change in A(Ni) of --0.1\,dex
decrease A(O) of 0.02\,dex.
The CN contribution to the complete EW in the 636\pun{nm} line is smaller
(12.5\%).

\subsection{HD 30562}

\begin{figure}
\resizebox{\hsize}{!}{\includegraphics[clip=true,angle=0]{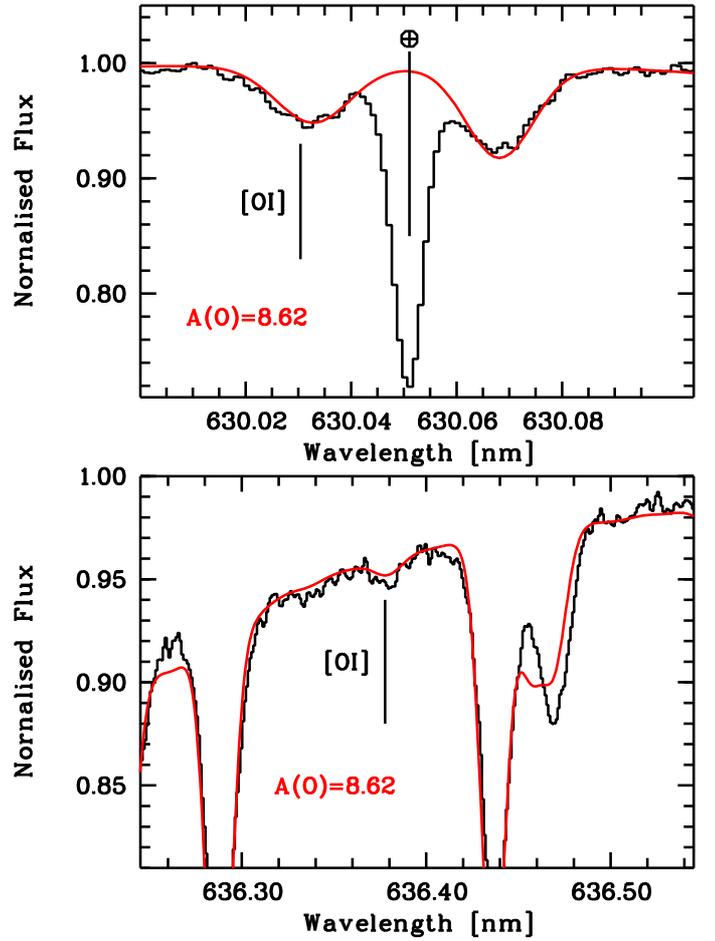}}
\caption{HARPS observed spectrum of HD\,30562 (solid black) is overimposed on
the synthetic spectrum, computed with A(O)=8.62\,dex (solid red) for the
the 630\pun{nm} and the 636\pun{nm} lines.
}
\label{hd30562}
\end{figure}

We analysed the HARPS spectra for HD\,30562, but unfortunately one of the telluric
lines of the range falls on the wing of the 630.0\,nm [OI] line.
We computed a synthetic spectrum ATLAS+SYNTHE with the stellar parameters
from \citet{feltzing98} (5876/4.0/+0.19), and [O/Fe]=--0.6.
The quality of the data is not good enough to permit a line profile fitting. 
Synthetic spectra are, however, shown in Fig.\ref{hd30562}.
At this uncertainty, the two lines seem to agree, but the telluric
line absorption close to the 630\,nm line probably alters the profile
of the oxygen line. As a result A(O)=8.62 can be seen as
an upper limit for the line at 630.0\,nm, while it is perhaps
a lower limit for the line at 636.3\,nm.

\subsection{$\alpha$\,Centauri\,A}

\begin{figure}
\resizebox{\hsize}{!}{\includegraphics[clip=true,angle=0]{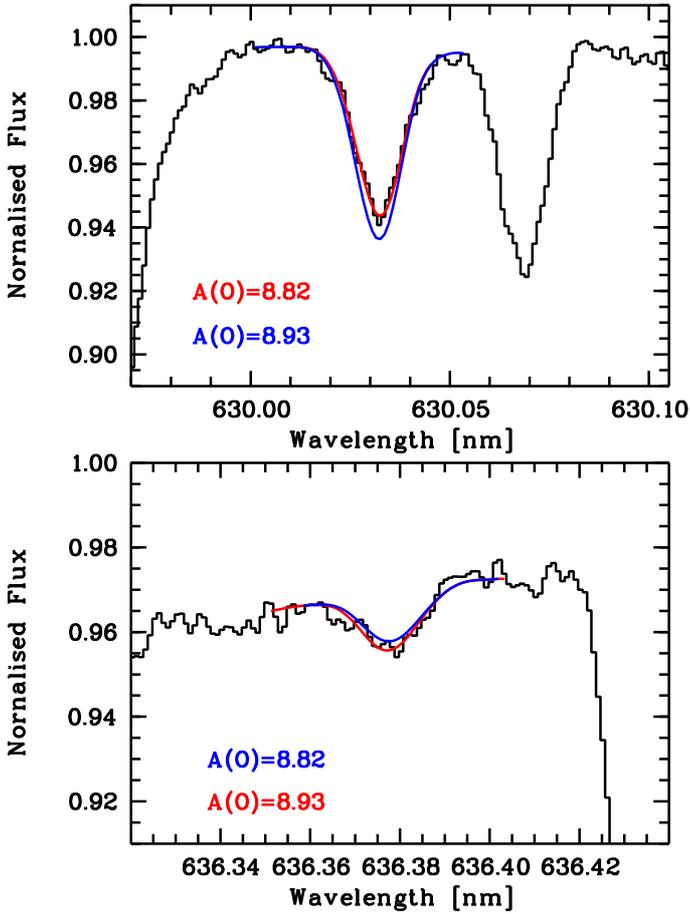}}
\caption{HARPS observed spectrum of $\alpha$\,Centauri A (solid black) 
overimposed on two synthetic spectra computed with A(O)=8.82 (solid red)
and A(O)=8.93 (solid blue), the best fit for the resonance and the subordinate
line, respectively.
}
\label{acen}
\end{figure}

We downloaded HARPS observations from the ESO 
archive\footnote{http://archive.eso.org/eso/eso\_archive\_adp.html}.
The stellar parameters (5824/4.34/+0.24) are from \citet{portodemelo08}.
We used an ATLAS12 model computed with the parameters of the star.
The two [OI] lines are in strong disagreement, giving an oxygen abundance
of A(O)=8.82 from the 630\,nm line and A(O)=8.92 from the line at 636.3\,nm 
(see the fits in Fig.\,\ref{acen} and the results in Table\,\ref{results}).
From the figure it can appear that the fit of the subordinate line is suboptimal
because the shape of the \ion{Ca}{i} auto-ionisation line is not properly modelled.

An increase of the Ni abundance of 0.1\,dex reduces the abundance
of oxygen by 0.06\,dex. This value is not sufficient to reconcile
the oxygen derived from the two [OI] lines.
When only O and Ni are included in the fit of the 630\,nm line,
the fit of the line profile gives A(O)=8.86 and 8.92 if A(Ni)=6.49 and 6.39, respectively.

According to a SYNTHE simulation, when an input ATLAS model is used,
the Ni contribution to the complete EW of the 630\pun{nm} blend is
large (48.6\% in EW) and
the CN contribution to the complete EW in the 636\pun{nm} line is
also large, but smaller than Ni in the resonance line
(29.0\%).

\subsection{$\alpha$\,Centauri\,B}

We downloaded HARPS observations from the ESO archive.
The stellar parameters (5223/4.44/+0.25) are from \citet{portodemelo08}.
We used an ATLAS12 model computed with the parameters of the star.
The two [OI] lines are in strong disagreement,
but we think that it is not possible to derive
A(O) from the 636\,nm line. In fact, in this blend the major contribution
is due to CN molecules, and by increasing A(O) the total EW is reduced.
Also, the contribution of Ni in the blend is dominant; in fact, a decrease 
of A(Ni) by 0.1\,dex introduces an increase in A(O) by 0.20\,dex.

According to a SYNTHE simulation, when an input ATLAS model is used,
the Ni contribution to the complete EW of the 630\pun{nm} blend is
large (69.2\% in EW) and
the CN contribution to the complete EW in the 636\pun{nm} line is
also large, but smaller than Ni in the resonance line
(62.8\%).
For the subordinate line a change in A(O) changes the strengths of
the lines of CN in the opposite direction.

\subsection{Procyon}

We also considered the spectrum of Procyon, retrieved from the UVES-POP database \citep{uvespop}
and adopted the parameters
of 6500\pun{K}/4.0/0.0 from \citet{Steffen}.
The 636\pun{nm} line implies a higher A(O) than the 630\pun{nm} line, but this line is
very weak and the S/N of the spectrum does not permit a definite determination
of A(O) from this line.

\begin{figure}
\resizebox{\hsize}{!}{\includegraphics[clip=true,angle=0]{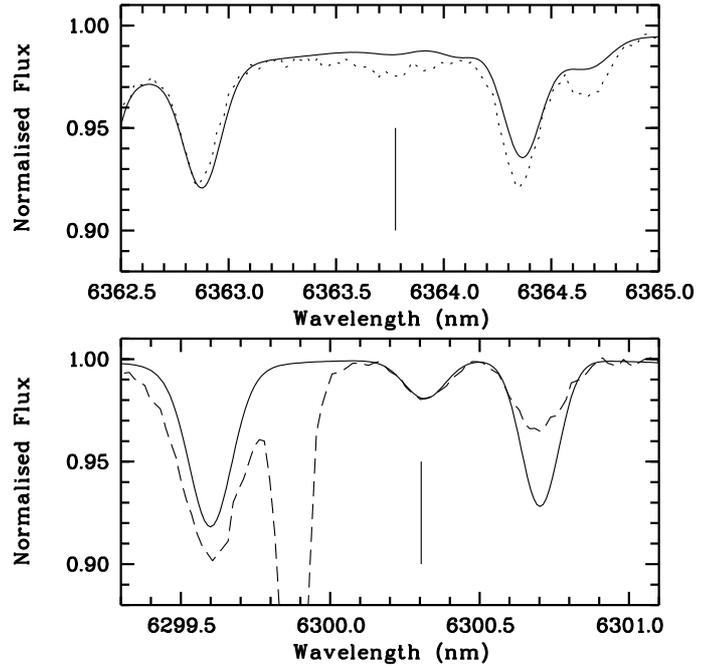}}
\caption{Observed flux spectrum of Procyon
showing the 630\pun{nm} line (dashed) and the 636\pun{nm} line (dotted) 
overimposed on an ATLAS+SYNTHE synthetic spectrum (A(O)=8.53, A(Ni)=6.25 
solid line). The vertical lines indicate the location of the forbidden O lines.
}
\label{procyon}
\end{figure}

From Fig.\,\ref{procyon} it is
clear that by adopting A(Ni)=6.25 the oxygen abundance implied by the 630\pun{nm}
line is at least 0.2\pun{dex} {\em lower} than the solar abundance; this
low-oxygen abundance fails to reproduce the 636\pun{nm} line. This could indicate
that the contribution of \ion{Ni}{i} to the 630\pun{nm} line is, as in the
Sun, overestimated. But it could also be that the same unidentified 
blend in the 636\pun{nm} line in the solar spectrum is present in Procyon.

\section{Uncertainties}

Our goal is the comparison of the oxygen abundance from
the two [OI] lines. We think we can neglect any uncertainty
of the oscillator strength of the oxygen lines because
they come from the same source and we can expect the same
effect on both lines.
We will also neglect the uncertainties related to effective
temperature and gravity. The two oxygen lines are similarly
sensitive to changes in temperature and gravity, so any effect
would not affect the difference of the oxygen abundance derived from these two lines.
This is not true for their blends, Ni and the CN lines, which can have a different
reaction to a change in \teff\ and \glog.
But the stars we analyse are bright so that stellar parameters
should be well known. The effect due to uncertainties
on \teff\ and \glog\ on Ni are not large: a change in 100\,K
and 0.5\,dex, respectively, results in a change of about 
10\% and 6\% in the EW of Ni, respectively,
in a model with parameters close to solar. 
The CN lines are not so sensitive to gravity 
(0.5\,dex change produces a difference in the EW of about 7\%),
while their EW change by about a factor 2 for a change in temperature of 100\,K.

We also neglect any uncertainty due to deviation from the assumed
local thermal equilibrium (LTE) and to granulation effects (3D corrections).
We know that the two levels involved in the [OI] transitions are close
to LTE conditions. But this is not true for the lines blending the oxygen features.
No study on NLTE is available on Ni transitions or on CN lines,
but we do not expect this weak line to deviate strongly from LTE conditions.
The 3D corrections were investigated in the solar case \citep{oxy}, and we know
from our experience that they are small for weak lines in models
close to solar metallicity.

To synthesise the CN lines we used the line lists from Kurucz.
We compared the synthesis with the one obtained by using the line lists 
kindly provided by Bertrand Plez.
For the solar model we found a difference in EW of about 8\% and 2\% in the CN
blending the 630 and 636\,nm [OI] lines. For the latter, although the
difference in EW is negligible, the shape of the CN lines is different.

\subsection{[OI] 630\,nm line}

The uncertainty of the \loggf\ of the Ni blending component is 0.06\,dex.
The uncertainty of the Ni abundance depends on the star.
Since the stars we are analysing are bright, we can expect
an uncertainty of about 0.05\,dex. 
Taking into account both the uncertainty of the \loggf\ and of the A(Ni),
we can suppose an uncertainty of about 0.1\,dex for the Ni line.
This is translated into an uncertainty of about 0.07\,dex in the oxygen abundance
for $\alpha$\,Cen\,A
and 0.11\,dex for $\alpha$\,Cen\,B.
For giants the effect of the Ni uncertainty on the A(O) determination is smaller
(e.g. less than 0.02\,dex for Arcturus and Pollux).
The other blending lines usually have a smaller effect.
For the giants, completely neglecting them gives a difference of 
about 0.03\,dex in the A(O) determination.
For a dwarf star ($\alpha$\,Cen\,A) there is an increase in A(O) of 0.04\,dex.
The effect on $\alpha$\,Cen\,B is much larger, 0.09\,dex.
If taken into account in a cool dwarf, the weaker blends on the 630\,nm line
can introduce an uncertainty up to about 0.05\,dex.
The uncertainty on A(O) due to continuum placement is about 0.03\,dex in dwarfs,
while in giants the effect is small and can be neglected.

Considering an average value,
the total uncertainty on the A(O) determination can be estimated to be
about 0.10\,dex for dwarf stars and about 0.04\,dex for giant stars.

\subsection{[OI] 636\,nm line}

Any uncertainty in the shape of the \ion{Ca}{i} auto-ionisation line 
and the Ca abundance introduces an uncertainty of the A(O) abundance 
derived from the [OI] 636\,nm line. This problem affects only
dwarf stars; the \ion{Ca}{i} auto-ionisation line is not visible
in the spectra of giant stars. 
As stated in Sec.\,\ref{sun}, an increase of the Ca abundance of 0.1\,dex
produces an increase of 0.02\,dex of A(O).
Since the [OI] line is weak and occupies only a small
range on the wing of the \ion{Ca}{i} auto-ionisation line, we do not
think that the shape of this line can have a major effect on A(O).

Another effect is the uncertainty in wavelength and \loggf\ of the CN lines.
An increase of 0.2 on the \loggf\ of the CN lines produces a decrease
of 0.06\,dex in the oxygen abundance in the solar case.
We think that the Sun can be a representative general case; only $\alpha$\,Cen\,A and B
have a larger contribution to the EW due to CN lines.

We tested the two CN line lists in the case of Pollux.
The line list provided by B. Plez introduces an increase in A(O) of 0.07\,dex.
While in $\alpha$\,Cen\,A, the two line lists give a difference of less than 0.01\,dex.

The abundances of C and N also affect the A(O) derived from the 636\,nm line,
but in different ways for hot and cool stars.
An increase of 0.1\,dex of A(C) and A(N), at fixed A(O), increases by about 4-6\%
the EW of the blend O+CN (CN contributes about 14\%) in a the solar spectrum, 
while it slightly decreases (by about 2\%) the EW of the blend O+CN 
(CN contributes of about 12\%) in the spectrum of Arcturus.
This decrease in the giant stars is related to the CO formation.
These effects are anyway small.

The impact on the A(O) determination due to the continuum placement is about 0.03\,dex.
The total uncertainty of the A(O) determination can be estimated to be
about 0.15\,dex for dwarf stars and about 0.10\,dex for giant ones.

\section{Discussion and conclusions}

\begin{figure}
\resizebox{\hsize}{!}{\includegraphics[clip=true,angle=0]{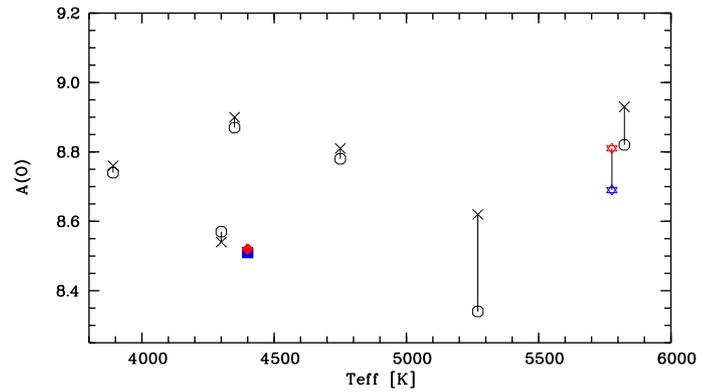}}
\caption{Oxygen abundances derived from the [OI] line at 630\,nm (circles)
and the line at 636\,nm (crosses) as a function of the effective temperature.
For plotting purposes the abundances of HD\,26297 are upshifted by 1\,dex and
represented as filled blue squares and red lozenges, respectively.
The solar values are indicated as blue and red stars.
}
\label{t_o}
\end{figure}

We compared the two forbidden oxygen lines in the Sun, 
five giants, one sub-giant, and four dwarfs.
The results are summarised in Fig.\,\ref{t_o}.
For the giants the oxygen abundance derived from the two lines agree within errors.
For two dwarf stars we have no
conclusive evidence: 
in Procyon the 636\pun{nm} line is too weak to
provide a  measurement, although the subordinate line
definitely seems to be too strong in comparison to the other; 
the spectrum of HD\,30562 is contaminated by a telluric absorption
on the 630.0\,nm line. 
For the sub-giant Capella and the two dwarfs ($\alpha$\,Centauri A and B)
the disagreement of the A(O) determination from the two [OI] lines is clear.

We verified that the strength of the CN lines is a function of the oxygen abundance:
the larger A(O), the weaker are the CN lines lying in the range
of the 636\,nm [OI] line.
This happens because the increase in the number of atoms of oxygen
increases the CO formation at the expense of CN molecules.
This effect is evident in giants and dwarfs, but it is larger in giants.
Due to this effect, the blend of O+CN decreases in giants by increasing
A(C) and A(N) at fixed oxygen abundance 
(4.6\% and 2.4\% decreases in the EW of the blend O+CN for an increase
of A(C) and A(N) of 0.2\,dex and 0.1\,dex, respectively), 
while in dwarfs the increase in A(C) and A(N) results
in a larger EW for the O+CN blend.

The reason for the 
discrepancy in the Sun that we previously favoured \citep{oxy} was that the CN molecules
blending the subordinate line are poorly known.
The idea was that the \loggf\ values underestimate the contribution of CN to
the blend or some CN lines are completely missing, resulting in a too
high estimated contribution from oxygen.
This view is in disagreement with the use of the recent CN line compilation of Plez,
which gives a smaller contribution of CN in the blend.
We also verified in this work that for cold giant stars,
where the contribution from CN molecules is not negligible, the two [OI] lines
give consistent oxygen abundances.
Since in the giant stars the Ni contribution to the EW of the 
O+Ni feature at 630\,nm 
becomes smaller than the contribution of CN
in the 636\pun{nm} line, we conclude that the CN components in the
subordinate line are probably not responsible for the disagreement of the two lines
in dwarfs.

Three possibilities are still in the game: i) the \loggf\ value of the \ion{Ni}{i}
line has an uncertainty larger that expected, ii) the shape of the \ion{Ca}{i} auto-ionisation line
influences the A(O) derived more than expected, iii) there is an unknown line in the
vicinity of the 630\,nm oxygen line, with atomic
characteristics to be strong enough in the atmospheres of dwarf stars.

It is difficult to believe that the \loggf\ value of the \ion{Ni}{i}
line blending the 630\,nm line is in error; the value is from \citet{Johansson}, 
who did a very careful investigation
and claim an uncertainty of 14\%, which is not sufficient to solve the problem.
But there is strong support in the sense that the \loggf\ is overestimated.
The study of solar spectra at different limb angles supports a Ni \loggf\ about 0.1\,dex lower.

We introduced a weak (\loggf =--2.974) \ion{Fe}{i} line at 636.378\,nm, of high-excitation energy 
(${\rm E_{\rm low}}=5.50$\,eV) so that is negligible for a giant star, but it has a non-negligible
contribution in a dwarf star.
A line (at 636.4763\,nm) with these parameters exists in the line list of Kurucz.
The introduction of such line has no impact on the A(O) derived from the 636\,nm line
in Arcturus. For Pollux we see a decrease in A(O) of 0.02\,dex, for $\alpha$\,Centauri\,A
of about 0.07\,dex, and for the Sun of 0.05\,dex.
An opportune high-excitation \ion{Fe}{i} line is a possible explanation.

The profile of the \ion{Ca}{i} auto-ionisation line is not properly modelled.
Although synthetic spectra are not able to reproduce satisfactorily the
\ion{Ca}{i} auto-ionisation line, on whose red wing
the 636\pun{nm} is formed, this cannot 
be the reason for the disagreement of the
two lines among dwarf stars. A better
modelling of the \ion{Ca}{i} auto-ionisation line   
would result in larger A(O) abundances derived from the
subordinate line.

Our results can be summarised as follows.
\begin{enumerate}
\item
In giants the two forbidden lines provide a consistent oxygen abundance
to within 0.1\pun{dex}.
\item
CN blending in the subordinate line is minor
and does not prevent an accurate A(O) determination from this line
in giants.
\item
In dwarfs and sub-giants, A(O) derived from the two [OI] lines is not in agreement.
\item
The profile of the \ion{Ca}{i} auto-ionisation line, visible in dwarfs and sub-giants,
is not properly modelled by synthetic spectra.
\item
An increase in the strength of the \ion{Ca}{i} auto-ionisation line by changing the
calcium abundance improves the agreement between the observed solar profile and the
synthetic spectrum in the range of the [OI] 636\,nm line. However,it does not have a
strong effect on the oxygen abundance. 
A better reproduction of the \ion{Ca}{i} auto-ionisation
line would provide even larger oxygen abundances from this
line, thus increasing the discrepancy.
\item
An increase of 0.2\,dex in the oscillator strengths of the CN lines 
is not sufficient to make the oxygen abundances from the two [OI] lines match.
\item
The use of an updated line list for the CN lines does not solve the problem.
\end{enumerate}

This evidence strongly suggests that either
the \loggf\ of the nickel line is smaller
than measured by \citet{Johansson} 
or that an unknown blend at 636\pun{nm} affects the spectrum of dwarfs.
All other possible explanations for the discrepancy are ruled out.

To make progress  in the investigation of the two [OI] lines,
major effort should be made 
to check for possible blends present only in dwarf star spectra
and the \loggf\ on the \ion{Ni}{i} line should be investigated again.
A theoretical investigation
towards a better reproduction of the \ion{Ca}{i}
auto-ionisation line
is also highly desirable, although it cannot solve
the discrepancy.


\begin{acknowledgements}
Based on data obtained from the ESO Science Archive Facility.
PB acknowledges support from the Programme National
de Physique Stellaire (PNPS) and the Programme National
de Cosmologie et Galaxies (PNCG) of the Institut National de Sciences
de l'Univers of CNRS.  
EC and HGL acknowledge financial support
by the Sonderforschungsbereich SFB881 ``The Milky Way
System'' (subprojects A4 and A5) of the German Research Foundation
(DFG).
\end{acknowledgements}

\bibliographystyle{aa}

\end{document}